\documentclass[12pt, table]{article}

\usepackage{fancyhdr}
\usepackage[table,xcdraw]{xcolor}
\usepackage{natbib}

\usepackage[document]{ragged2e}
\usepackage{ulem}
\usepackage{graphics}
\usepackage[export]{adjustbox}
\usepackage{epsfig}
\usepackage{wrapfig}
\usepackage{shadow}
\usepackage{color}
\usepackage{dcolumn}	
\usepackage{amsmath}
\usepackage{amsfonts}
\usepackage{amssymb}
\usepackage{multirow}
\usepackage{enumitem}
\usepackage{graphicx}
\usepackage{booktabs}
\usepackage{lipsum}
\usepackage{siunitx}
\usepackage{gensymb}
\usepackage[final]{pdfpages}
\usepackage{pdflscape}
\usepackage{upgreek}
\usepackage[hidelinks]{hyperref}
\usepackage[many]{tcolorbox}
\usepackage[labelfont=bf,skip=0pt]{caption}
\usepackage{titlesec}

\usepackage{authblk}
\usepackage{orcidlink}

\titlespacing*{\section}
{0pt}{0.5\baselineskip}{0.5\baselineskip}
\titlespacing*{\subsection}
{0pt}{0.5\baselineskip}{0.5\baselineskip}
\titlespacing*{\subsubsection}
{0pt}{0.5\baselineskip}{0.5\baselineskip}

\definecolor{myred}{HTML}{FB6542}
\definecolor{dodgerblue}{RGB}{30, 144, 255}
\definecolor{myblue}{HTML}{375E97}
\definecolor{tableblue}{HTML}{1A73C9}
\definecolor{mygrey}{HTML}{363636}

\DeclareCaptionFont{myblue}{\color{myblue}}
\usepackage{sectsty}
\sectionfont{\color{myblue}}
\subsectionfont{\color{myblue}}

\newcommand{\Rs}{\(\mathsf{R}_\odot\)}

\newtcolorbox{bangbox}[2][]
{
  enhanced,
  colframe=white,
  title={#2},
  width=\textwidth, 
  colback=tableblue,
  coltext=white,
  center title,
  colbacktitle=myred,
  coltitle=white,
  fonttitle=\bfseries
}

\newtcolorbox{bangboxnh}[1][]
{
  enhanced,
  colframe=white,
  width=\textwidth, 
  colback=tableblue,
  coltext=white} 

\newcommand{\swri}
	{Southwest Research Institute, Boulder, CO}
\newcommand{\ncar}
	{National Center for Atmospheric Research, Boulder, CO}
\newcommand{\predsci}
    {Predictive Science Inc., San Diego, CA}
\newcommand{\umn}
    {University of Minnesota, Minneapolis, MN}
\newcommand{\amu}
    {American University, Washington, DC}
 \newcommand{\msfc}
    {NASA Marshall Space Flight Center, Huntsville, AL}
\newcommand{\cfa}
    {Center for Astrophysics $|$ Harvard \& Smithsonian, Cambridge, MA}
\newcommand{\calb}
    {University of California, Berkeley, CA}
\newcommand{\gsfc}
    {NASA Goddard Space Flight Center, Greenbelt, MD}

\usepackage[margin=1.0in, headheight=30pt]{geometry}
\begin{document}

\widowpenalty=0
\clubpenalty=0
\flushbottom

\makeatletter
\renewcommand\Authfont{\fontsize{12}{14.4}\selectfont}
\renewcommand\Affilfont{\fontsize{9}{10.8}}
\renewcommand\AB@affilsepx{, \protect\Affilfont}
\makeatother


\title{\textbf{Improving Multi-Dimensional Data Formats, Access, and Assimilation Tools for the Twenty-First Century}}
\date{\vspace{-5ex}} 


\author[1]{\textbf{Primary Author:} Daniel~B.~Seaton%
\orcidlink{0000-0002-0494-2025}}
\author[1]{\textbf{Core Team Co-Authors:} Amir~Caspi%
\orcidlink{0000-0001-8702-8273}}
\author[2]{Roberto~Casini%
\orcidlink{0000-0001-6990-513X}}
\author[3]{Cooper~Downs%
\orcidlink{0000-0003-1759-4354}}
\author[2]{Sarah~E.~Gibson%
\orcidlink{0000-0001-9831-2640}}
\author[2]{Holly~Gilbert%
\orcidlink{0000-0002-9985-7260}}
\author[4]{Lindsay~Glesener%
\orcidlink{0000-0001-7092-2703}}
\author[5]{Silvina~Guidoni%
\orcidlink{0000-0003-1439-4218}}
\author[1]{J.~Marcus~Hughes%
\orcidlink{0000-0003-3410-7650}}
\author[6]{David~McKenzie%
\orcidlink{0000-0002-9921-7757}}
\author[1]{Joseph~Plowman%
\orcidlink{0000-0001-7016-7226}}
\author[7]{Katharine~K.~Reeves%
\orcidlink{0000-0002-6903-6832}}
\author[8]{Pascal~Saint-Hilaire%
\orcidlink{0000-0002-8283-4556}}
\author[9]{Albert~Y.~Shih%
\orcidlink{0000-0001-6874-2594}}
\author[1]{Matthew~J.~West%
\orcidlink{0000-0002-0631-2393}}

\affil[1]{\swri}
\affil[2]{\ncar}
\affil[3]{\predsci}
\affil[4]{\umn}
\affil[5]{\amu}
\affil[6]{\msfc}
\affil[7]{\cfa}
\affil[8]{\calb}
\affil[9]{\gsfc}
\maketitle
\vspace{-2ex}
\centering
\textbf{Additional Co-Authors:} Refer to attached spreadsheet.

\justifying

\begin{bangbox}{Synopsis}
\justifying
\noindent
Heliophysics image data largely relies on a forty-year-old ecosystem built on the venerable Flexible Image Transport System (FITS) data standard. While many \textit{in situ} measurements use newer standards, they are difficult to integrate with multiple data streams required to develop global understanding. Additionally, most data users still engage with data in much the same way as they did decades ago. However, contemporary missions and models require much more complex support for 3D multi-parameter data, robust data assimilation strategies, and integration of multiple individual data streams required to derive complete physical characterizations of the Sun and Heliospheric plasma environment. In this white paper we highlight some of the 21$^\mathsf{st}$~century challenges for data frameworks in heliophysics, consider an illustrative case study, and make recommendations for important steps the field can take to modernize its data products and data usage models.
Our specific recommendations include:
\vspace{-1ex}
\begin{itemize}[leftmargin=*]
    \item Investing in data assimilation capability to drive advanced data-constrained models, \vspace{-1.ex}
    \item Investing in new strategies for integrating data across multiple instruments to realize measurements that cannot be produced from single observations, \vspace{-1.ex}
    \item Rethinking old data use paradigms to improve user access, develop deep understanding, and decrease barrier to entry for new datasets, \vspace{-1.ex}
    \item Investing in research on data formats better suited for multi-dimensional data and cloud-based computing.
\end{itemize}
\end{bangbox}

\fancypagestyle{firstpage}{%
  \chead{ 
  \begin{bangboxnh}
    \centering
    \textbf{A White Paper Submitted to the Solar \& Space Physics (Heliophysics) Decadal Survey --- 2024--2033} 
  \end{bangboxnh}
} }
\thispagestyle{firstpage}


\newpage
\cfoot{{\color{myblue} \thepage}}

\setcounter{page}{1}
\section{The Challenge for Heliophysics Data Standards}
\label{sec:intro}
\justifying

The Flexible Image Transport System (FITS), the prevailing standard for image-based data products in heliophysics, was developed in the late 1970s, when computerized analysis of electronic image data was just becoming the norm in astronomy. FITS was intended to stop the proliferation of ad hoc formats and provide a simple standard -- well suited to storage on magnetic tape -- that was both human and machine readable \citep{Wells1981}.

The standard's design was brilliant: engineered simultaneously to be appropriately sized and formatted for tape files, compatible with almost all computers of the era, and accessible even to data users who were unfamiliar with the new format. In so doing, the standard also (preemptively) complied with recommended best practices for archival data formats, which mandate such formats should still be interpretable even in the absence of the computer or software systems for which they were designed \citep{NAP4871}.

Since those early days, FITS has been fully standardized, upgraded, and refined to better carry complex, compressed, multi-dimensional data needed for modern observations and computers, reaching version 4.0 in the present day \citep{FITS4}. FITS has also become the single standard for nearly all heliophysics image data, and the community has developed infrastructure that allows users to search and download data \citep[i.e., the Virtual Solar Observatory][]{Hill2009} in FITS format from dozens of instruments.

Though FITS is capable of containing multi-dimensional data, its primary purpose at conception was to transmit image data, generally serially, in individual files, for local calibration and analysis. Historically the data use model for heliophysics observations in FITS format has been to obtain individual observations in individual files, transport them to a local file system, and calibrate and analyze them there.

In an era in which data were generally 2D representation of camera output, whether image or spectra, this model was generally workable. However, many contemporary and proposed instruments produce vastly more complex data \citep[e.g.,][]{Cheung2019, Golub2020}, require advanced image processing tools and significant computing resources to calibrate and process \citep[e.g.,][]{Winebarger2019, DeForest2017}, and produce time-varying, multi-dimensional datasets that are difficult to represent under old paradigms. Increasing data product volume has made local analysis of advanced data less and less feasible, even in spite of improvements in internet bandwidth, necessitating online analysis environments with simplified or degraded representations of quantitative data \citep{Muller2017}. Many data products are not optimized for integration with physical models, reducing their value as model drivers or constraints, and slowing progress on data assimilation in heliophysics compared to other data-rich scientific domains. Integrating data products across multiple instruments is often infeasible, or requires extensive expertise in multiple projects, which few researchers possess. 
\vspace{-1ex}

\begin{bangboxnh}
\justifying
Since the advent of FITS 40 years ago, heliophysics has evolved from a data-limited research environment with nascent numerical modeling capabilities, to a data-rich one with advanced models with extensive data assimilation needs. There is a critical need for new approaches to data products and data assimilation strategies. Through the development of the \textit{COMPLETE} mission concept \citep[see additional white papers by][]{CaspiWP2002_Mission, CaspiWP2002_Science}, we have identified new strategies for complex, multi-perspective, multi-dimensional data products, and recommendations for investments that could help realize a new vision for heliophysics data for the next century. In this white paper we highlight a case study (Sec.~\ref{sec:case_study}) and provide recommendations for these improvements and investments (Sec.~\ref{sec:recs}).
\end{bangboxnh}
\vspace{-1ex}

\section{Case Study: Data Integration for COMPLETE}
\label{sec:case_study}

The COMPLETE mission concept embodies the need for advanced data products and processing for contemporary missions, where multiple data streams from disparate measurements must be integrated within a unified physical framework. COMPLETE comprises two integrated instrument suites -- a comprehensive 3D magnetograph and broadband spectral imager -- distributed across multiple spacecraft at differing solar view angles. The comprehensive magnetograph combines surface field measurements from a photospheric magnetograph instrument with magnetic diagnostics in the corona using a Hanle-effect Lyman-$\alpha$ polarization coronagraph \citep{Raouafi2016b}. The broadband spectral imager combines observations from $\gamma$-rays, X-rays, and EUV along with multi-messenger energetic neutral atom observations to deduce plasma properties in the corona. To integrate these disparate measurements, the instruments are specifically curated and co-optimized to produce mutually compatible observations, which then must be assimilated into an overall 3D data framework within physical context.

Some plasma properties, such as temperature and density, can be deduced from broadband spectral images straightforwardly, using differential emission measure techniques \citep[DEM; e.g.,][]{Cheung2015, Plowman2020}. However, traditional applications of these techniques can only reveal the properties projected into 2D images, and integrated along the line of sight.

Techniques to invert Stokes profiles to determine magnetic fields have existed for decades \citep{Auer1977} and are now both robust and computationally advanced \citep{Borrero2011}. However, to characterize the corona's magnetic properties, these techniques must be coupled to models that extrapolate or predict the coronal magnetic field. In the absence of additional constraints, solving for the coronal field is a massively underdetermined problem \citep{Plowman2021}. However, Hanle-effect measurements can provide the missing constraint -- if they can be assimilated into the overall data/model framework.

Assimilation of multi-instrument, multi-messenger, multi-dimensional data is part of routine
\begin{wrapfigure}{r}{0.56\columnwidth}
\centering
\includegraphics[width=0.55\columnwidth]{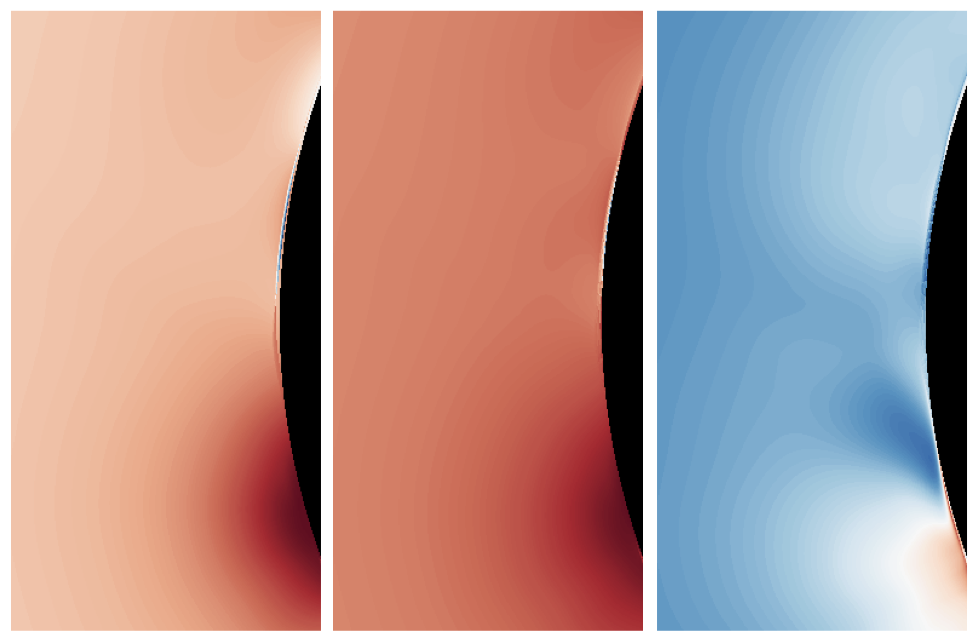}
\caption{\small \justifying Forward-modeled Hanle-effect polarization through different line-of-sight depths (0.1, 0.2, \& 1.0\,\Rs, left to right), demonstrating the diagnostic potential of these observations of coronal fields. The underlying model is a Magnetohydrodynamic Algorithm outside a Sphere (MAS) simulation of the corona at the time of the 2017 total solar eclipse \citep{Mikic2017}.
 } 
\label{fig:hanle}
\end{wrapfigure}
\noindent research and forecasting operations in data-rich fields such as atmospheric science and meteorology \citep[see the review by][and references therein]{Lahoz2014}. While this is new territory for heliophysics, with appropriate investment to adapt existing strategies there is no reason heliophysics cannot achieve the same level of sophistication and success as these other fields. For example, for COMPLETE, we studied a straightforward but powerful method that can integrate magnetic field information and plasma properties to generate a 3D model of key physical parameters in the corona using multi-perspective observations. These reconstructed parameters can then serve as model initial conditions or constraints within more sophisticated simulations.

\begin{wrapfigure}{r}{0.56\columnwidth}
\centering
\includegraphics[width=0.54\columnwidth]{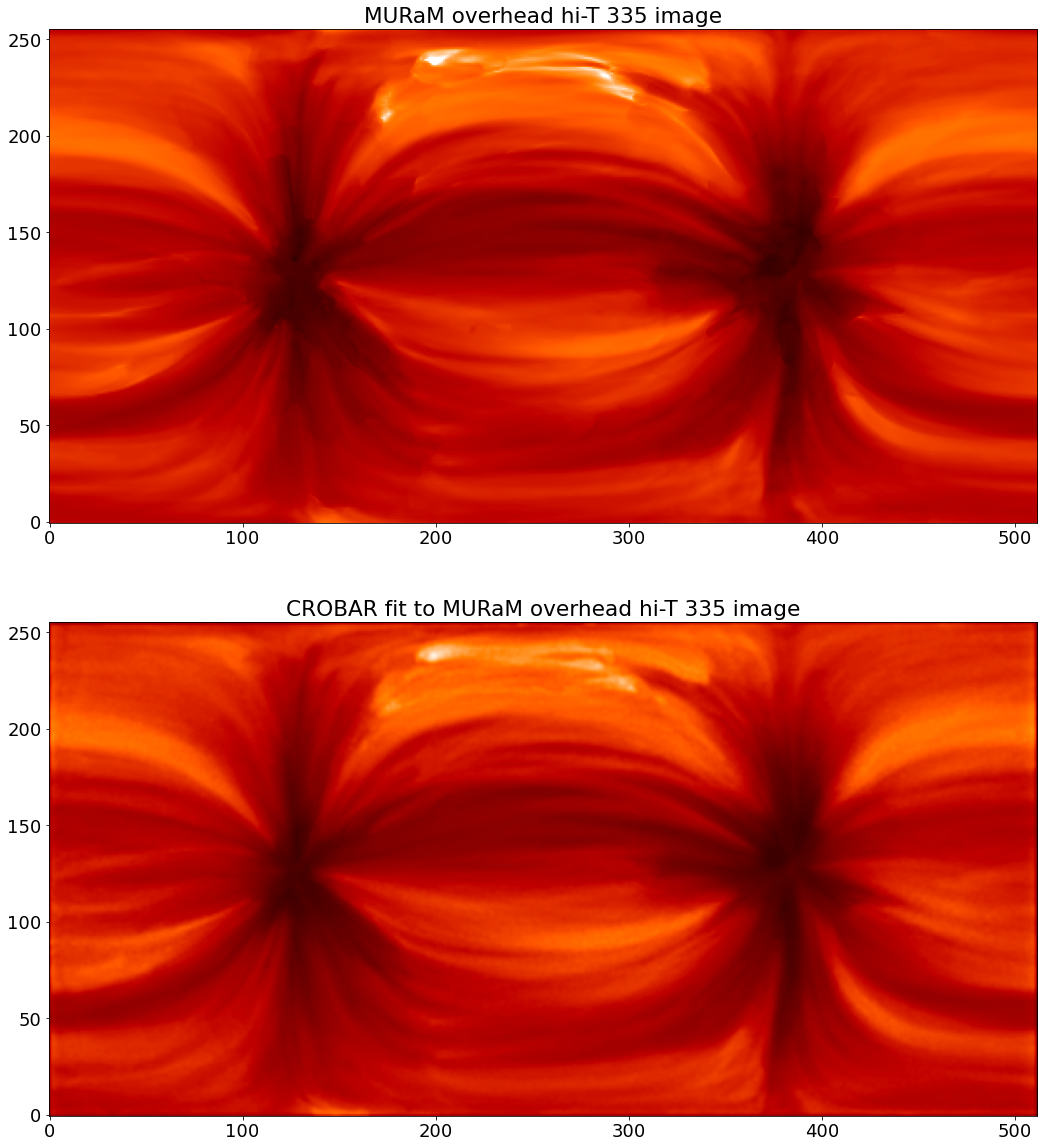}
\caption{\small \justifying Synthesized image of coronal loops from the MuRAM model \citep[top;][]{Rempel2017} compared to an image derived from a 3D CROBAR reconstruction \citep[bottom;][]{Plowman2022}.}
\label{fig:crobar_3D}
\end{wrapfigure}

\subsection{Reconstructing the 3D Corona}
\label{ssec:reconstructions}

The approach we are developing for COMPLETE uses an extension of the Coronal Reconstruction Onto B-Aligned Regions \citep[CROBAR;][]{Plowman2021, Plowman2022} method. CROBAR leverages the fact that the high conductivity of the corona confines plasma to magnetic field lines (the so-called ``frozen-flux'' condition) to generate 3D reconstructions of plasma distribution within a magnetic extrapolation from photospheric boundary conditions. By coupling the reconstruction technique to data-constrained magnetic extrapolations, DEM tools, and multiple perspectives, it is possible to obtain an accurate 3D reconstruction of the coronal temperature, density, pressure, and magnetic field -- and therefore derived and correlated properties like plasma $\beta$ -- within the reconstructed volume.

CROBAR presently uses a linear force-free magnetic model tuned by comparison with optically thin emission (e.g., EUV) images. However, its reconstructions can also drive forward models of Hanle effect observables \citep[][see Fig.~\ref{fig:hanle}]{Gibson2016}, which are directly connected to the coronal magnetic field, in much the same way that CROBAR optimizes the magnetic field by comparison with the emission observations. The method already achieves impressive fidelity with EUV observations alone (Fig.~\ref{fig:crobar_3D}), and with direct coronal field constraints it will be better still.

\begin{figure}
\centering
\includegraphics[width=0.98\columnwidth]{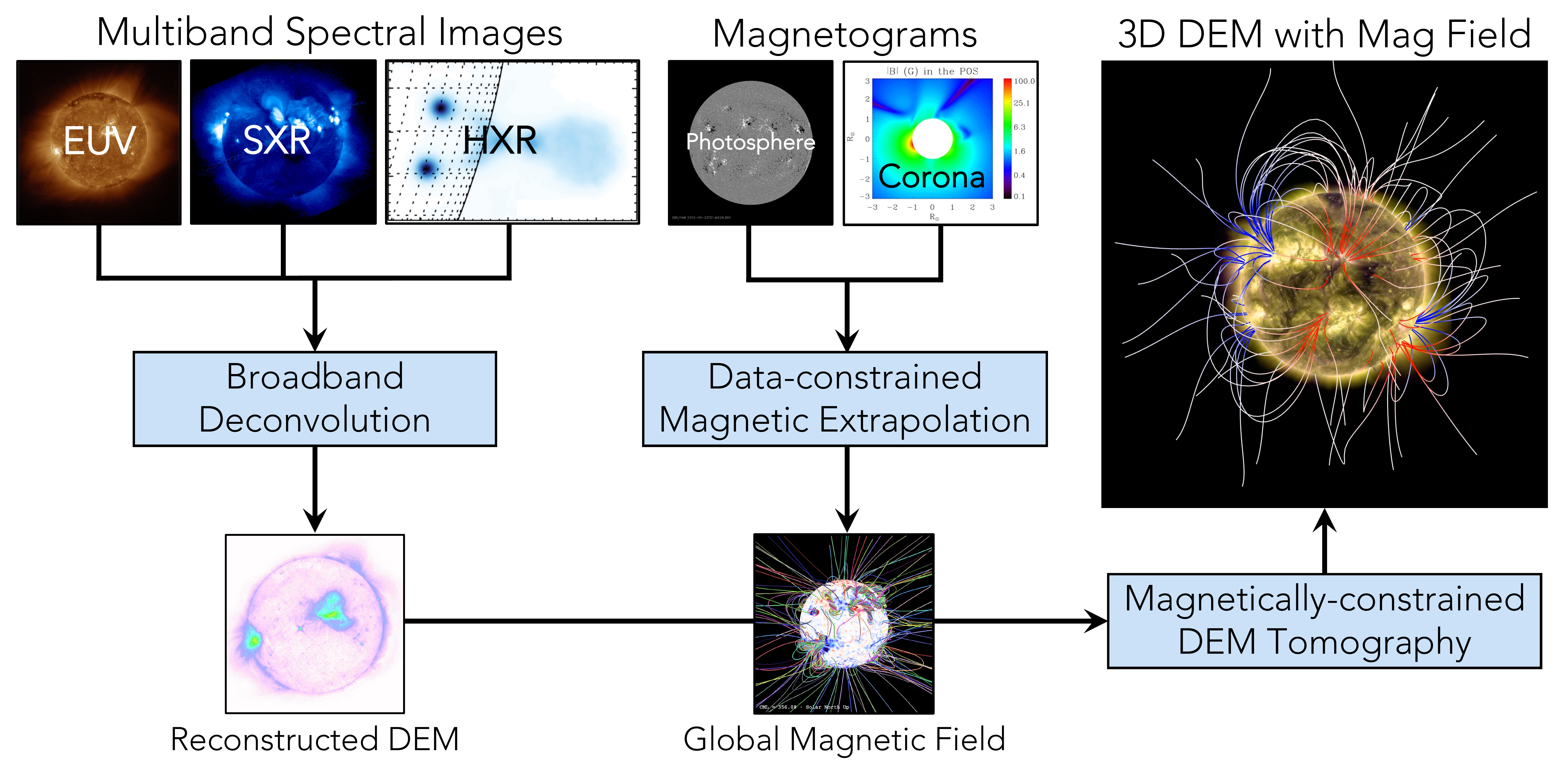}
\caption{\small \justifying COMPLETE's strategy to integrate multi-faceted data into a unified data-constrained model of conditions in the corona highlights the types of opportunities advanced data assimilation and model strategies may provide. 
 \vspace{-3ex}} 
\label{fig:flow}
\end{figure}

Such a reconstruction technique (see Fig.~\ref{fig:flow}), which synthesizes multi-point observations of surface magnetic fields, coronal magnetic field diagnostics, and broadband spectral images, demonstrates, in a simple package, the feasibility of reconstructing key coronal parameters with limited computational overhead. {\color{myblue}\textbf{Reconstructed 3D parameters could then serve as data-driven initial conditions or other constraints for numerical simulations of the corona, and thus provide a straightforward path for a fully integrated data assimilation strategy}}.

\subsection{A Challenge to Traditional Data Management Strategies}
\label{ssec:challenge}

3D reconstructions of many physical parameters, such as this example, present a major opportunity for heliophysics, but also a major challenge. As envisioned by the COMPLETE concept, each reconstruction represents a snapshot in time with a set of high-resolution, multi-dimensional data: three spatial dimensions in spherical coordinates, vector magnetic field, and plasma parameters including temperature, density, and pressure (which themselves may need to be multidimensional, as coronal plasma can be multi-thermal) as well as the original observables. These reconstructions must also include extensive metadata to describe their complex primary data so that researchers and their models and analysis tools can interact with it appropriately.

It is unlikely such reconstructions can be efficiently represented within the current limitations of the venerable FITS standard, and new file formats will be needed. Potential candidate standards already exist in other data-rich fields, including: \vspace{-1ex}
\begin{itemize}
    \item Network Common Data Form \citep[netCDF;][]{netCDF}, widely used for atmospheric and in situ space physics data and optimized for multi-dimensional arrays with complex metadata; \vspace{-4ex}
    \item Environmental Systems Research Institute (ESRI) Shapefiles \citep{ESRI_SF}, widely used for vector data in for geographic information system applications; \vspace{-1.5ex}
    \item Zarr \citep{zarr}, a young, promising cloud-optimized format, which has seen some use in heliophysics through recent projects at NASA's Frotier Development Lab in 2022. \vspace{-2ex}
\end{itemize}
\vspace{1ex}

Each of these has seen use in applications with similar requirements to those of modern heliophysics data, and offers a degree of standardization and community adoption (albeit, within other communities) that could help to facilitate adoption within heliophysics as well. However, more research is required to develop specific requirements necessary to identify a common format that would be appropriate for advanced data such as ours across an entire community.

Likewise, existing analysis frameworks may be insufficient to quantitatively explore these new products. Though tools like jHelioviewer provide good pathfinders for \textit{qualitative} analysis of multi-perspective, multi-band image data, they do not permit deep exploration of quantitative data in 3D. The size of these multidimensional products is likely to upend traditional models of data distribution, potentially necessitating new cloud-based analysis environments, where users can work within the full multi-dimensional data set, extract key measurements, and only download the subset of measurements, plots, or renderings they require locally at the end of their analysis activity. A few such environments already exist, including example implementations using Google's \textit{Colab} notebook-based coding environment \citep{Tamayo2021} and the Space Radiation Intelligence System \citep[SPRINTS;][]{Engell2017} project, but much more research and development is required to realize the full-featured analysis environment that will be required for such modern data.

\section{Recommendations: Data for 21$^\mathsf{st}$ Century Science}
\label{sec:recs}

To continue to support cutting edge science during the coming decade, the heliophysics community must adopt new, innovative thinking about what data products are and how to use them. We have identified four major strategic priorities to address the needs of the data- and model-rich environment developing within the field.
\begin{bangbox}{Strategic Priorities}
\begin{itemize}[leftmargin=*]
    \item Invest in data assimilation capability to drive advanced data-constrained models, \vspace{-1.ex}
    \item Invest in new strategies for integrating data across multiple instruments to realize measurements that cannot be produced from single observations, \vspace{-1.ex}
    \item Rethink old data use paradigms to improve user access, develop deep understanding, and decrease barriers to entry for new datasets, \vspace{-1.ex}
    \item Invest in research on data formats better suited for multi-dimensional data and cloud-based computing.
\end{itemize}
\end{bangbox}

\noindent {\color{myblue}\textbf{Data Assimilation.}}
Many numerical models in heliophysics use direct measurements as boundary conditions for advanced calculations, but most of these models still lack sufficient constraints to ensure they accurately represent the underlying physical systems. However, with multi-perspective observations, routine measurements of coronal magnetic fields, and advanced data processing techniques, new constraints will become available in the coming decade. Novel approaches are required to ingest these new measurements, along with direct in situ sampling of coronal plasma from missions like Parker Solar Probe, into models that were not developed with such constraints in mind. Interdisciplinary research efforts, particularly in coordination with other fields, such as atmospheric science, with robust support for model data assimilation will be of value. These strategies, and the advanced models they support, will be of value both for basic research and space weather forecasting -- strategic planning will allow missions and data products to be developed with these broad applications in mind.\\

\noindent {\color{myblue} \textbf{Multi-observation Integration.}}
Traditional models for data products were developed when most data were single, 2D images or spectra, and were intended to be used in isolation from other observations. As more observational capability emerged, researchers developed strategies to deal with complementary observations (e.g., EUV coronal images from AIA and corresponding magnetographs from HMI). A few techniques, such as DEM analysis, exist that can ingest data from multiple observatories, but they are not robust, since the observations themselves were not developed with cross-instrument integration in mind.

New missions in development (such as the COMPLETE mission concept) prioritize this kind of cross-instrument data integration to achieve results that cannot be obtained from single observations alone. Both investments in data integration methods, such as those outlined in Sec.~\ref{sec:case_study} here, and strategic planning, to identify and develop compatible datasets for this purpose, are needed.

An potential pathfinder may be the Polarimeter to Unify the Heliosphere and Corona Small Explorer mission \citep[PUNCH;][]{deforest2022polarimeter}, which integrates the observations of four individual imagers into complete, polarimetric observations of the corona and heliosphere between 6--180\,\Rs. The instrument and data processing design strategies developed for PUNCH point to both challenges and opportunities to overcome them that will have broad applicability to further research into multi-instrument measurements. Future missions that make more complex observations -- and produce more complex data -- will require even more complex strategies. Targeted investments into research on multi-instrument, multi-perspective integrated data are required.
\\

\noindent {\color{myblue} \textbf{A New Data Use Paradigm.}}
Present-day data analysis models, in which data users acquire and analyze data locally, have their roots in the pre-internet era -- far before the development of today's cloud computing strategies. But as data have grown in complexity, this model has led to increased barriers for researchers looking to work with multiple observation sets, both because the data volume can become restrictively large, and because specialized expertise is required to work with new data.

New thinking about how to constitute data products, as well as how and where we work with them, can break down these barriers. Data products that contain derived physical measurements are easier for new users to interpret than direct observations (from which these parameters must be extracted), and do not require the user access complex and computationally expensive processing software to make measurements. Cloud computing solutions are increasingly viable, and permit users to access data without extensive transfers to local systems. Additional work is needed to improve cloud-based visualization capabilities, determine how to manage data egress costs, and develop robust, powerful analysis tools for these environments. 
\\

\noindent {\color{myblue} \textbf{Advanced Data Formats.}}
The new data applications discussed above may result in products or work environments for which current data standards are not optimized. Community-wide investments are required to identify new standards, or improvements to existing standards, to support more advanced, modern data applications. Standards at use in other communities may be applicable or adaptable to the heliophysics community's needs. Extensions of existing standards may also be possible, analogous to the incorporation of Zarr's cloud-ready capabilities into netCDF via the NCZarr format \citep[see ``NCZarr Introduction'' in][]{netCDF}. Community assessment of the requirements for contemporary data standards and investments to support the development (or re-development) of data formats to meet these requirements are critical.

Formats and analysis environments that leverage the cloud represent a major step forward, but also require community investments to ensure long-term maintenance, stewardship, and appropriate curation. Funding agencies must develop strategies to ensure sustained data access under such a model, as well as resources to support appropriate levels of data egress (e.g., for rendered movies, figures, and subsets of data required for local analysis).

\begin{bangboxnh}
\justifying
\noindent Investments in these strategic priorities, coupled with the development of new observational capabilities for 3D plasma parameters and magnetic fields, can provide keys to true \textit{transformative progress} within heliophysics during the coming decades. By building robust data products that leverage lessons learned from cross-discipline research on data assimilation, heliophysics can finally realize both comprehensive understanding of the Sun-Heliosphere system as a whole, and major gains in our ability to study, forecast, and track drivers of space weather that are critical to long-term risk management in our space-faring, technologically-driven society.
\end{bangboxnh}

\newpage

\setcounter{page}{1}
\cfoot{{\color{myblue} References -- \thepage}}

\bibliography{references.bib}
 \bibliographystyle{aasjournal}

\end{document}